\documentclass[runningheads]{llncs}
\usepackage{listings}
\usepackage{amsmath,amsfonts,amssymb}
\usepackage{tikz}
\usepackage{subfig}
\usepackage{array}
\usepackage{pifont}
\usepackage[nolist,nohyperlinks]{acronym}
\usepackage{multirow}
\usepackage{pgfplotstable}
\usepackage{longtable}

\usetikzlibrary{fit, spy, positioning, calc, arrows, arrows.meta,
  decorations.pathreplacing, decorations.markings, shapes}

\newcolumntype{M}[1]{>{\centering\arraybackslash}m{#1}}

\newcommand{\code}[1]{\texttt{#1}}%

\usepackage{refcount}
\newcounter{nalg}

\renewcommand{\thenalg}{\arabic{nalg}} 
\DeclareCaptionLabelFormat{algocaption}{Algorithm \thenalg}

\lstnewenvironment{algorithm}[1][] 
                  {
    \setcounter{lstlisting}{\value{nalg}}
    \refstepcounter{nalg}
    \captionsetup{labelformat=algocaption, labelsep=colon}
    \lstset{ 
      mathescape=true,
       comment = [l]{//},
       frame=tb,
       numbers=left, 
       numberstyle=\scriptsize,
       basicstyle=\footnotesize, 
       keywordstyle=\bfseries\em,
       keywords={,input, min, output, return, solve, datatype, function, in, if, else, foreach, not, while, endwhile, begin, end, relax, }
       numbers=left,
       xleftmargin=.04\textwidth,
       #1 
    }
}
{}

\lstdefinestyle{mipsstyle}{%
  comment = [l]{\#},
  frame=tb,
  numbers=left,
  numberstyle=\scriptsize,
  basicstyle= \scriptsize,
  keywordstyle=\bfseries\em,
  keywords={,lw, sw, add, addi, addiu, jr, jal, nop, move, slti, beqz, blez, mul, }
  numbers=left,
  xleftmargin=.08\textwidth,
}

\lstdefinestyle{cstyle}{%
  language=c,
  basicstyle= \scriptsize,
  xleftmargin=.08\textwidth,
}

\lstdefinestyle{cfgstyle}{%
    comment = [l]{\#},
    basicstyle= \scriptsize,
    keywordstyle=\bfseries\em,
    keywords={,lw, sw, add, addi, ret_ra, addiu, jr, jal, copy, phi, mul, beq, retra, slti, b, blez, }
}

\newsavebox{\measurebox}

\makeatletter
\newcolumntype{"}{@{\hskip\tabcolsep\vrule width 1pt\hskip\tabcolsep}}
\makeatother

\begin{acronym}
\acro{CSP}{Constraint Satisfaction Problem}
\acro{COP}{Constraint Optimization Problem}
\acro{LNS}{Large Neighborhood Search}
\acro{RS}{Random Search}
\acro{ATGP}{Automatic Generation of Architectural Tests}
\acro{ROP}{Return-Oriented Programming}
\acro{ISA}{Instruction Set Architecture}
\acro{IoT}{Internet of Things}
\acro{JOP}{Jump-Oriented Programming}
\acro{JIT-ROP}{Just-in-time Re\-turn-Oriented Programming}
\acro{HD}{Hamming Distance}
\acro{RD}{Register Hamming Distance}
\acro{LD}{Levenshtein Distance}
\acro{CFI}{Control-Flow Integrity}
\acro{CFG}{Control-Flow Graph}
\acro{XoM}{Execute-Only Memory}
\acro{DivCon}{Diversity by Construction}
\acro{IR}{Intermediate Representation}
\acro{MIR}{Machine Intermediate Representation}
\acro{CP}{Constraint Programming}
\end{acronym}

\begin{document}
%


\title{Constraint-Based Software Diversification for Efficient Mitigation of Code-Reuse Attacks}

\titlerunning{Constraint-Based Software Diversification Against Code-Reuse Attacks}
%

\author{Rodothea Myrsini Tsoupidi\inst{1} \and
  Roberto Casta{\~{n}}eda Lozano\inst{2} \and
  Benoit Baudry\inst{1}
}
%
\institute{%
  KTH Royal Institute of Technology, Sweden\\
  \email{\{tsoupidi,baudry\}@kth.se} \and
  University of Edinburgh, United Kingdom\\
  \email{roberto.castaneda@ed.ac.uk}}

\maketitle              
\begin{abstract}
  Modern software deployment process produces software that is uniform, and hence
  vulnerable to large-scale code-reuse attacks.
  \emph{Com\-pil\-er-based diversification} improves the resilience and security of
  software systems by automatically generating different assembly code
  versions of a given program.
  Existing  techniques are efficient but do not
  have a precise control over the quality of the generated code variants.

  This paper introduces \emph{Diversity by Construction (DivCon)}, a
  constraint-based compiler approach to soft\-ware diversification.
  Unlike previous approaches, DivCon allows users to control and adjust the
  conflicting goals of diversity and code quality.
  A key enabler is the use of Large Neighborhood Search (LNS) to generate highly
  diverse assembly code efficiently.

  Experiments using two popular compiler benchmark suites confirm that there is
  a trade-off between quality of each assembly code version and diversity of the entire pool of versions. Our results show that
  DivCon allows users to trade between these two properties by generating diverse assembly
  code for a range of quality bounds.
  In particular, the experiments show that DivCon is able to mitigate code-reuse
  attacks effectively while delivering near-optimal code ($<10\%$ optimality
  gap).

  For constraint programming researchers and practitioners, this paper
  demonstrates that LNS is a valuable technique for finding diverse solutions.
  For security researchers and software 
  engineers, DivCon extends the scope of
  compiler-based diversification to performance-critical and resource-constrained
  applications. 
\end{abstract}

\renewcommand{\thelstlisting}{\arabic{lstlisting}}

\pgfplotstableset{
  begin table=\begin{longtable},
  end table=\end{longtable},
}

\section{Introduction} 
\label{sec:introduction}

Good software development practices, such as code reuse~\cite{Krueger1992},
continuous deployment, and automatic updates contribute
to the emergence of software monocultures~\cite{birman2009monoculture}.
While such monocultures facilitate software distribution, bug reporting,
and software authentication,
they also introduce serious risks related to the wide spreading of attacks
against all users that run identical software.

\textit{Software diversification} is a method to mitigate the problems caused by uniformity.
Similarly to biodiversity,
software diversification improves the resilience and security
of a software system~\cite{baudryMultipleFacetsSoftware2015} by introducing diversity in it.
Software diversification
can be applied in different phases of
the software development cycle, i.e.\
during implementation, compilation,
loading, execution, and more~\cite{larsen_sok_2014}.
This paper is concerned with \emph{compiler-based} diversification,
which automatically generates different assembly code versions from a single
source program.

Modern compilers do not merely aim to generate correct code, but also code that
is of high quality.
Existing compiler-based diversification techniques are efficient and effective
at diversifying assembly code~\cite{larsen_sok_2014} but do not have a precise
control over its quality and may produce unsatisfactory results.
These techniques (discussed in Section~\ref{sec:rel}) are either based on
randomizing heuristics or in high-level superoptimization methods that do not
capture accurately the quality of the generated code.

This paper introduces \ac{DivCon}, a compiler-based diversification approach
that allows users to control and adjust the conflicting goals of quality of each
code version and diversity among all versions.
\ac{DivCon} uses a \ac{CP}-based compiler backend to generate multiple solutions
corresponding to functionally equivalent program variants
according to an accurate code quality model.
The backend models the input program, the hardware architecture, and the
compiler transformations as a constraint problem, whose solution corresponds to
assembly code for the input program.

The use of \ac{CP} makes it possible to 1) control the quality of the generated
solutions by constraining the objective function, 2) introduce
application-specific constraints that restrict the diversified solutions, and 3)
apply sophisticated search procedures that are particularly suitable for
diversification.
In particular, \ac{DivCon} uses \ac{LNS}~\cite{Shaw1998}, a popular
metaheuristic in multiple application domains, to generate highly diverse
solutions efficiently.

Our experiments compiling 17 functions from two popular compiler benchmark
suites to the MIPS32 architecture confirm that there is a trade-off between code
quality and diversity, and demonstrate that \ac{DivCon} allows users to navigate
this conflict by generating diverse assembly code for a range of quality bounds.
In particular, the experiments show that \ac{DivCon} is able to mitigate
code-reuse attacks effectively while guaranteeing a code quality of 10\% within
optimality.

For constraint programming researchers and practitioners, this paper
demonstrates that LNS is a valuable technique for finding diverse solutions.
For security researchers and software engineers, DivCon extends the scope of
compiler-based diversification to performance-critical and resource-constrained
applications, and provides a solid step towards secure-by-construction software.

\paragraph{Contributions.}
To summarize, this paper:
\begin{itemize}
\item proposes a \ac{CP}-based technique for compiler-based, quality-aware
  software diversification (Section~\ref{sec:approach});
\item shows that \ac{LNS} is a promising technique for generating highly diverse
  solutions efficiently (Section~\ref{ssec:evallns});
\item quantifies the trade-off between code quality and diversity (Section~\ref{ssec:gap_eval}); and
\item demonstrates that \ac{DivCon} mitigates code-reuse attacks effectively
  while preserving high code quality (Section~\ref{ssec:sec_eval}).
\end{itemize}


\section{Background}
\label{sec:background}

This section describes code-reuse attacks (Section~\ref{ssec:attacks}),
diversification approaches in \ac{CP} (Section~\ref{ssec:bakdivcp}), and
combinatorial compiler backends (Section~\ref{ssec:unison}).

\subsection{Code-reuse Attacks}
\label{ssec:attacks}

Code-reuse attacks take advantage of
memory vulnerabilities, such as buffer overflows,
to reuse 
program code for malicious purposes.
More specifically, code-reuse attacks
insert data into the program memory
to affect the control flow of the program and
execute code that is
valid but unintended.

\ac{JOP}\footnote{This paper focuses on \ac{JOP} due to the characteristics
  of MIPS32, but could be generalized to other code-reuse attacks such as
  \ac{ROP}~\cite{shacham_geometry_2007}.}  is a code-reuse attack~\cite{checkoway_return-oriented_2010,bletsch_jump-oriented_2011} that combines
different code snippets from the original program code to form a Turing complete
language for attackers.  These code snippets terminate with a branch
instruction.
The building blocks of a \ac{JOP} attack are \emph{gadgets}:
 meta-instructions that consist of
one or multiple code snippets and have
specific semantics.
Figure~\ref{lst:mips}
shows a \ac{JOP} gadget found by the \emph{ROPgadget} tool~\cite{ROPGadget2020}
in a MIPS32 binary.
Assuming that the attacker controls the stack,
lines 2 and 3
load attacker data in registers \$s2 and \$s4, respectively.
Then, 
line 4 jumps to the address
of register \$t9.
The last instruction (line 5) is placed in a delay slot and, hence, it is executed
before the jump~\cite{Sweetman2006}.
The semantics of this gadget depends on the actual attack payload and
might be to load a value to register \$s2 and \$s4.
Then, the program jumps to the next gadget that resides
at the stack address of \$t9.

\begin{figure}[t]
  \subfloat[Original gadget.]{%
    \label{lst:mips}
    \begin{minipage}[b]{.48\textwidth}
\begin{lstlisting}[style=mipsstyle]
0x9d001408: ...
0x9d00140c: lw    $s2, 4($sp)
0x9d001410: lw    $s4, 0($sp)
0x9d001414: jr    $t9
0x9d001418: addiu $sp, $sp, 16

\end{lstlisting}
      \vspace{-0.2cm}
    \end{minipage}
}
\subfloat[Diversified gadget.]{%
    \label{lst:mips2}
    \begin{minipage}[b]{.48\textwidth}
\begin{lstlisting}[style=mipsstyle]
0x9d001408: lw    $s2, 4($sp)
0x9d00140c: nop
0x9d001410: lw    $s4, 0($sp)
0x9d001414: jr    $t9
0x9d001418: addiu $sp, $sp, 16
\end{lstlisting}
      \vspace{-0.2cm}
    \end{minipage}
}
\caption{\label{fig:mips_example} Example gadget diversification in MIPS32 assembly code}
\end{figure}
  
Statically designed \ac{JOP} attacks use the absolute
binary addresses for installing the
attack payload.
Hence, a simple change in the instruction schedule of the program
as in Figure~\ref{lst:mips2} prevents a \ac{JOP}
attack designed for Figure~\ref{lst:mips}.
An attacker that designs an attack based on the binary of
the original program
assumes the presence
of a gadget (Figure~\ref{lst:mips})
at position
0x9d00140c.
However, in the diversified
version, address 0x9d00140c
does not start with the initial \texttt{lw} instruction
of Figure~\ref{lst:mips}, and by the end of
the execution of the gadget, register \$s2 does not
contain the attacker data.
In this way, diversification can break the semantics of the gadget
and mitigate an attack against the diversified code.

\subsection{Diversity in \acl{CP}}
\label{ssec:bakdivcp}

While typical \ac{CP} applications aim to discover either some solution or the
optimal solution, some applications require finding \textit{diverse} solutions
for various purposes.

Hebrard \emph{et al.}\ \cite{hebrard_finding_2005} introduce
the \textsc{MaxDiverse$k$Set} problem, which consists in finding
the most diverse set of $k$ solutions,
and propose an exact and an incremental algorithm for solving it.
The exact algorithm does not scale to a large number of solutions
\cite{van_hentenryck_constraint-based_2009,ingmar_modelling_2020}.
The incremental algorithm selects solutions iteratively by solving
a distance maximization problem.

\ac{ATGP} is an application of \ac{CP} that
 requires
generating many diverse solutions.
Van Hentenryck \emph{et al.}\ \cite{van_hentenryck_constraint-based_2009}
model  \ac{ATGP}
as a \textsc{MaxDiverse$k$Set} problem and solve it
using the incremental algorithm of Hebrard et.\ al.
Due to the large number of
diverse solutions required (50-100), 
Van Hentenryck \emph{et al.}\
replace the maximization step
with local search.

In software diversity, solution quality is of paramount importance.
In general, earlier \ac{CP} approaches to diversity are concerned with satisfiability only.
An exception is the approach of Petit \emph{et al.}~\cite{petit_finding_2015}.
This approach modifies the objective function for assessing both solution
quality and solution diversity, but does not scale to the large number of
solutions required by software diversity.
Ingmar \emph{et al.}~\cite{ingmar_modelling_2020} propose a
generic framework for modeling diversity in \ac{CP}. For tackling the
quality-diversity trade-off, they propose constraining the objective function
with the optimal (or best known) cost \textit{o}.
\ac{DivCon} applies this approach by allowing solutions $p\%$ worse
than \textit{o}, where $p$ is configurable.

\subsection{Compiler Optimization as a Combinatorial Problem } 
\label{ssec:unison}

A \acf{CSP} is a problem specification $P =\langle V,U, C\rangle$,
where $V$ are the problem variables,
$U$ is the domain of the variables, and $C$ the constraints among the variables.
A \acf{COP}, $P =\langle V,U, C, O \rangle$, consists of a \ac{CSP}
and an objective function $O$. The goal of a \ac{COP} is to find a solution
that optimizes $O$.

Compilers are programs that generate low-level assembly code, typically
optimized for \textit{speed} or \textit{size}, from higher-level source code.
A compilation process can be modeled as a \ac{COP} by letting $V$ be the
decisions taken during the translation, $C$ be the constraints imposed by the
program semantics and the hardware resources, and $O$ be the cost of the
generated code.

\begin{figure} 
  \centering
\input{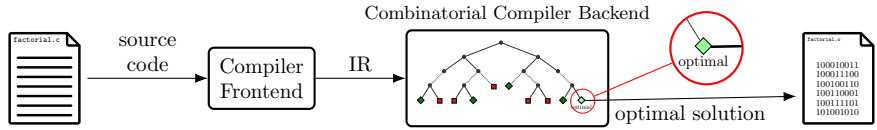}
\caption{\label{fig:unison}High-level view of a combinatorial compiler backend}
\end{figure}

Compiler backends generate low-level assembly code from an \ac{IR}, a program
representation that is independent of both the source and the target language.
Figure~\ref{fig:unison} shows the high-level view of a \emph{combinatorial}
compiler backend.
A combinatorial compiler backend takes as input the \ac{IR} of a program,
generates and solves a \ac{COP}, and outputs the optimized low-level assembly
code described by the solution to the \ac{COP}.

This paper assumes that programs at the \ac{IR} level are represented by
their \ac{CFG}.
A \ac{CFG} is a representation of the possible execution paths of a program,
where each node corresponds to a \emph{basic block} and edges correspond to
intra-block jumps.
A \emph{basic block}, in its turn, is a set of abstract instructions (hereafter
just \emph{instructions}) with no branches besides the end of the block.
Each instruction is associated with a set of operands characterizing its input
and output data.
Typical decision variables $V$ of a combinatorial compiler backend are the issue
cycle $c_i \in \mathbb{N}_0$ of each instruction $i$, the processor instruction
$m_i \in \mathbb{N}_0$ that implements each instruction $i$, and the processor
register $r_o \in \mathbb{N}_0$ assigned to each operand $o$.

\ac{DivCon} aims at  mitigating code-reuse attacks.
Therefore,
\ac{DivCon} considers the order of the
instructions
in the final binary,
which directly affects the feasibility of code-reuse attacks
(see Figures~\ref{lst:mips} and \ref{lst:mips2}). 
For this reason, the diversification model uses the issue cycle sequence
of instructions, $c = \{c_0,c_1,..., c_n\}$,
to characterize the diversity among different solutions.

\begin{figure}[t]
  \newcommand{\dollar}{\mbox{\textdollar}}
  \centering
  \subfloat[\label{fig:fact} C code]{
    \begin{tikzpicture}[scale=1, transform shape,
    nst2/.style={
      inner xsep = -3pt,
      inner ysep = -6pt,
      align = left,
      minimum width = 4cm,
      text width = 4cm,
      anchor = south},
    nst/.style={
      draw,
      rounded corners=2pt,
      inner xsep = -3pt,
      inner ysep = -6pt,
      align = left,
      minimum width = 3.5cm,
      text width = 3.5cm,
      anchor = south}]

\node[nst2] (func1) {
  \begin{lstlisting}[style=cstyle]
int factorial(int n) {
\end{lstlisting}};

\node[nst, below=0.1cm of func1] (b0) {
  \begin{lstlisting}[mathescape=true, style=cstyle]
int f;
f = 1;
\end{lstlisting}};

\node[nst, below=0.1cm of b0] (b1) {
  \begin{lstlisting}[mathescape=true, style=cstyle]
while(n > 0) {
  f *= n--;
}
\end{lstlisting}};

\node[nst, below=0.1cm of b1] (b2) {
  \begin{lstlisting}[mathescape=true, style=cstyle]
return f;
\end{lstlisting}};

\node[nst2, below=0.1cm of b2] (func2) {
  \begin{lstlisting}[ style=cstyle]
}
\end{lstlisting}};

\end{tikzpicture}
  }
  \hfill
  \subfloat[\label{fig:factcfg} \ac{IR}]{
    \begin{tikzpicture}[scale=0.9, transform shape,
  nst/.style={
    draw,
    rounded corners=2pt,
    inner xsep = 3pt,
    inner ysep = -5pt,
    align = left,
    minimum width = 3cm,
    text width = 3cm,
    anchor = south}]

  \node[nst] (b0) {
      \begin{lstlisting}[mathescape=true, style=cfgstyle]
0: $t_1 \gets$ $\dollar$a0
1: $t_2 \gets$ 1
2: blez $t_1,$ bb.2
      \end{lstlisting}};

  \node[nst, below=0.3cm of b0] (b1) {
      \begin{lstlisting}[mathescape=true, style=cfgstyle]
3: $t_2 \gets$ mul $t_2,$ $t_1$
4: $t_3 \gets$ slti $t_1,$ 2
5: $t_1 \gets$ addi $t_1,$ -1
6: beq $t_3,$ %0$,$ bb.1
7: b bb.2
      \end{lstlisting}};

    \node[nst, below=0.3cm of b1] (b2) {
      \begin{lstlisting}[mathescape=true, style=cfgstyle]
8: $\dollar$v0 $\gets t_2$
9: jr $\dollar$ra
      \end{lstlisting}};

    \node[inner sep=0.4mm, anchor=south west] at (b0.north west) {\scriptsize{bb.0}};
    \node[inner sep=0.4mm, anchor=south west] at (b1.north west) {\scriptsize{bb.1}};
    \node[inner sep=0.4mm, anchor=south west] at (b2.north west) {\scriptsize{bb.2}};

    \draw[->, >={Latex}] (b0) -- (b1);
    \draw[->, >={Latex}] (b1) -- (b2);
    \draw[->, >={Latex}] (b0.south west) to [out=230, in=130]  (b2.north west);
    \draw[->, >={Latex}] (b1.south east) to [out=30, in=330]  (b1.north east);

\end{tikzpicture}
    
  }
  \hfill
  \subfloat[\label{fig:cyc} Issue cycles]{
    \begin{tikzpicture}[scale=0.6, transform shape,
  nst/.style={rectangle, minimum width = 0.6cm, minimum height = 0.6cm, draw, anchor=north west},
  ast/.style={decorate,decoration={brace,amplitude=7pt},xshift=0pt,yshift=-4pt},
  mst/.style={black, midway, right, xshift=14pt}]

  \node[rectangle,
    minimum width =  0.6cm,
    minimum height = 6cm,
    draw,
    anchor=north west,
    rounded corners=1pt,
    semithick] at (0,0) {};

  \foreach \x in {0,...,9} {
    \node[nst] (n\x) at (0,-0.6*\x) {$c_\x$}; 
  }

\draw [ast] (n0.north east) -- (n2.south east) node [mst] {\footnotesize bb.0};
\draw [ast] (n3.north east) -- (n7.south east) node [mst] {\footnotesize bb.1};
\draw [ast] (n8.north east) -- (n9.south east) node [mst] {\footnotesize bb.2};
\node [right = 40pt of n1] {};
\node [left = 20pt of n1] {};
  
\end{tikzpicture}
    
  }
  \caption{\label{fig:factorial} Factorial function example}
\end{figure}
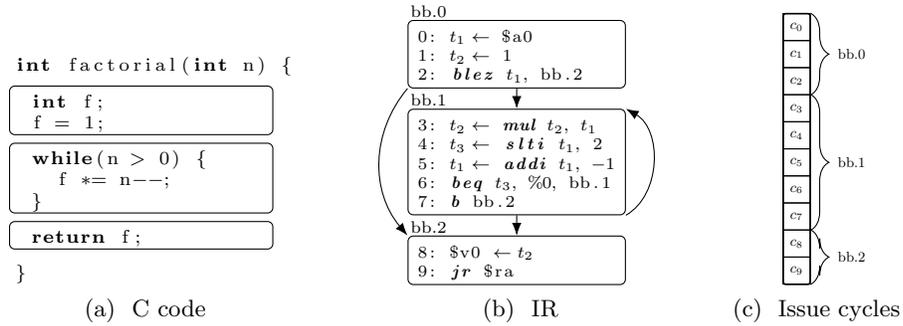

Figure~\ref{fig:fact} shows an implementation of the factorial
function in C where each basic block is highlighted.
Figure~\ref{fig:factcfg} shows the
\ac{IR} of the program.
The example \ac{IR} contains 10 instructions in three basic blocks: bb.0, bb.1,
and bb.2.
bb.0 corresponds to initializations, where \texttt{\$a0} holds the function argument
\texttt{n} and $t_1$ corresponds to variable \texttt{f}.
bb.1 computes the factorial in a loop by accumulating the result in $t_1$.
bb.2 stores the result to \texttt{\$v0} and returns.
Some instructions in the example are interdependent, which leads to
serialization of the instruction schedule.
For example, \textbf{\textit{beq}} (6) consumes data ($t_3$) defined by
\textbf{\textit{slti}} (4) and hence needs to be scheduled later.
Instruction dependencies limit the amount of possible assembly code versions and
can restrict diversity significantly, as seen in Section~\ref{ssec:evallns}.
Finally, Figure~\ref{fig:cyc} shows the arrangement of the issue cycle variables
in the constraint model used by the combinatorial compiler backend.

\section{DivCon}
\label{sec:approach}

\begin{figure}[t]
  \centering
  \input{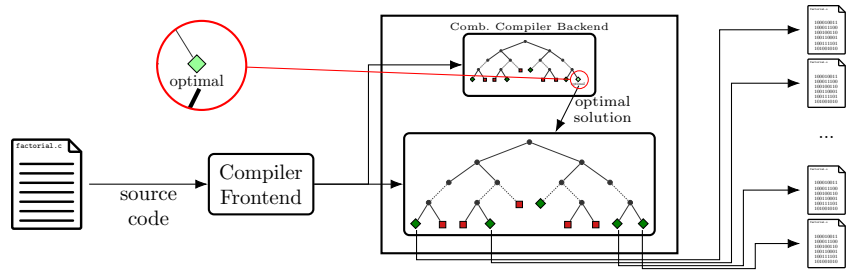}
  \caption{\label{fig:divcon} High-level view of \ac{DivCon}}
\end{figure}

This section introduces \ac{DivCon}, a software diversification method
that uses a combinatorial compiler backend to generate program variants.
Figure~\ref{fig:divcon} shows a high-level view 
of the diversification process.
\ac{DivCon} uses 1) the optimal solution to start the \textit{search} for
diversification and 2) the cost of the optimal solution to
restrict the variants within a maximum gap from the optimal.
Subsequently, \ac{DivCon} generates a
number of solutions to the \ac{CSP} that correspond
to diverse program variants.

The rest of this section describes the diversification
approach of \ac{DivCon}.
Section~\ref{ssec:problem}
formulates the diversification problem
in terms of the constraint
model of a combinatorial compiler backend,
Section~\ref{ssec:distances} defines  the 
distance measures, and finally, Section~\ref{ssec:search}
describes the search strategy for generating program variants.

\subsection{Problem Description}
\label{ssec:problem}
Let $P = \langle V,U,C\rangle$ be the compiler backend \ac{CSP} for the program
under compilation, $O$ be the objective function, and $o$ be the cost of the optimal or
best known solution to the \ac{COP}, $\langle V,U,C,O\rangle$.
Let $\delta$ be a function that
measures the distance between
two solutions of P (two such functions are defined in Section~\ref{ssec:distances}). 
Let $h \in \mathbb{N}$ be the minimum pairwise distance and $p \in \mathbb{R}_{\geq 0}$ be the maximum optimality gap specified by the user.
Our problem is to find a subset of the solutions to the \ac{CSP}, $S \subseteq sol(P)$,
such that
$\forall s_1, s_2 \in S\, . \, s_1 \neq s_2 
\implies \delta(s_1,s_2) \ge h$ and $\forall s \in S \, . \,
O(s) \leq (1+p) \cdot o $.

To solve the above problem, \ac{DivCon} employs the incremental algorithm listed in Algorithm~\ref{alg:search}.
Starting with the optimal solution $y_{opt}$,
the algorithm adds the distance constraint for $y_{opt}$ and the optimality constraint with $o = y_{opt}(O)$ (line 2).
Notation $\delta(y)$
is used instead of
$\delta(y, s)  \,|\, \forall s \in sol(\langle V,U,C'\rangle)$ for readability.
While the termination condition is not fulfilled (line 3), the algorithm uses
\ac{LNS} as described in Section~\ref{ssec:search} to find the next solution $y$ (line 4),
adds the next solution to the solution set $S$ (line 5),
and updates the distance constraints based on the latest solution (line 6).
When the termination condition is satisfied, the algorithm returns the set of solutions $S$
corresponding to diversified assembly code variants.

\begin{algorithm}[caption={Incremental algorithm for generating diverse solutions},
    label={alg:search}]
  S $\gets$ {$y_{opt}$}, y $\gets y_{opt}$,
  $C' \gets C \,\cup \{\delta(y_{opt}) \ge h$, $O(V) \leq (1 + p)\cdot o\}$
  while not term_cond() // e.g. |S| > k $\lor$ time_limit()
      y $\gets$ solve$_{LNS}$(relax(y), $\langle V,U, C'\rangle$)
      S $\gets$ S $\cup$ {y}
      $C' \gets C'\,\cup \{\delta(y) \ge h\}$
\end{algorithm}

Figure~\ref{fig:mips_variants} shows two MIPS32 variants of the
factorial example (Figure~\ref{fig:factorial}), which correspond to two
solutions of \ac{DivCon}.
The variants differ in two aspects: first, the \textbf{\textit{beqz}}
instruction is issued one cycle later in Figure~\ref{lst:fact2} than
in Figure~\ref{lst:fact1}, and second, the temporary variable $t_3$
(see Figure~\ref{fig:factorial}) is assigned to different MIPS32 registers (\$t0 and \$t1).

\begin{figure}[b]
  \subfloat[Variant 1.]{%
    \label{lst:fact1}
  \begin{minipage}[t]{0.38\textwidth} %
\begin{lstlisting}[style=mipsstyle, firstline=20,lastline=27]
bb.0: blez  $a0, bb.2
      addiu $v0, $zero, 1
bb.1: mul   $v0, $v0, $a0
      slti  $t0, $a0, 2
      beqz  $t0, bb.1
      addi  $a0, $a0, -1
bb.2: jr    $ra
      nop
\end{lstlisting}
    \vspace{-0.2cm}
    \
  \end{minipage}}
  \hfill
  \subfloat[Variant 2.]{%
    \label{lst:fact2}
  \begin{minipage}[t]{0.38\textwidth}
\begin{lstlisting}[style=mipsstyle, firstline=20,lastline=28]
bb.0: blez  $a0, bb.2
      addiu $v0, $zero, 1
bb.1: mul   $v0, $v0, $a0
      slti  $t1, $a0, 2
      nop
      beqz  $t1, bb.1
      addi  $a0, $a0, -1
bb.2: jr    $ra
      nop
\end{lstlisting}
    \vspace{-0.2cm}
  \end{minipage}}
  \caption{\label{fig:mips_variants} Two MIPS32 variants of the
factorial example in Figure~\ref{fig:factorial}}
\end{figure}

\subsection{Distance Measures}
\label{ssec:distances}
This section
defines two alternative 
distance measures:
\acf{HD}
and \acf{LD}.
Both distances operate on the schedule of the instructions, i.e.\
the order in which the instructions are issued in the CPU.

\paragraph{\acf{HD}.} \ac{HD}
is the Hamming distance~\cite{hamming1950error} between the
issue cycle variables of two  solutions.
Given two solutions $s, s' \in sol(P)$:
\begin{equation}
  \delta_{HD} (s,s') = \displaystyle\sum_{i=0}^{n} (s(c_i) \neq s'(c_i)),
  \label{eq:hdist}
\end{equation}
\noindent
where $n$ is the maximum number of instructions.

Consider Figure~\ref{lst:mips2}, a diversified version of the gadget
in Figure~\ref{lst:mips}.
The only instruction that differs from
Figure~\ref{lst:mips} is the instruction at line 1 that is
issued one cycle before. The two examples
have a \ac{HD} of one,
which in this case is enough for breaking
the functionality of the original gadget (see Section~\ref{ssec:attacks}).

\paragraph{\acf{LD}.}
\ac{LD} (or edit distance) measures the minimum number of edits, i.e.\
insertions, deletions, and replacements, that are necessary
for transforming one instruction schedule to another.
Compared to \ac{HD}, 
which considers only \textit{replacements},
\ac{LD} also considers \textit{insertions} and \textit{deletions}.
To understand this effect, consider Figure~\ref{fig:mips_variants}.
The two gadgets differ only by one \texttt{nop}
operation but \ac{HD} gives a distance
of three, whereas \ac{LD} gives one,
which is more accurate.
\ac{LD} takes ordered vectors as input, and thus requires an ordered
representation (as opposed to a detailed schedule) of the instructions.
Therefore, \ac{LD} uses vector $c^{-1}= channel(c)$, a sequence of instructions
ordered by their issue cycle.
Given two solutions $s, s' \in sol(P)$:
\begin{equation}
  \delta_{LD} (s,s') = \texttt{levenshtein\_distance}(s(c^{-1}),s'(c^{-1})), 
  \label{eq:levdist}
\end{equation}
\noindent
where \texttt{levenshtein\_distance} is the Wagner–Fischer algorithm~\cite{wagner1974string}
with time complexity 
$O(nm)$, where $n$ and $m$ are the lengths of the two sequences.

\subsection{Search}
\label{ssec:search}

Unlike previous CP approaches to diversity, \ac{DivCon} employs \acf{LNS} for diversification.
\ac{LNS} is a metaheuristic that defines a neighborhood, in which 
\textit{search} looks for better solutions, or, in our case, different solutions.
The definition of the neighborhood is through a \textit{destroy}
and a \textit{repair} function.
The \textit{destroy} function unassigns a subset of the variables in a given solution
and the \textit{repair} function finds a new solution by assigning
new values to the \emph{destroyed} variables.

In \ac{DivCon},
the algorithm starts with the optimal solution of the combinatorial
compiler backend.
Subsequently, it destroys
a part of the variables
 and continues with
the model's branching strategy
to find the next solution, applying a restart after a given number of failures.
\ac{LNS} uses the concept of \textit{neighborhood}, i.e.\ the variables that
\ac{LNS} may destroy at every restart.
To improve  diversity, the neighborhood for \ac{DivCon} consists of
all decision variables, i.e.\  the issue cycles $c$, the instruction implementations $m$,
and the registers $r$.
Furthermore, \ac{LNS} depends
on a \textit{branching strategy} to
guide the \textit{repair} search.
To improve security and allow \ac{LNS} to select diverse paths after every
restart, \ac{DivCon} employs a random variable-value selection branching
strategy as described in Table~\ref{tab:random}.
\begin{table}[b]
  \caption{\label{tab:randomcloriginal} \textsc{Original} and \textsc{Random} branching strategies}
  \centering
  \subfloat[\label{tab:cloriginal} \textsc{Original} branching strategy.]{
    {\footnotesize
      \begin{tabular}{|c|>{\centering}m{2cm}|c|}
        \hline
        Variable & Var. Selection & Value Selection \\\hline 
        $c_i$ &  in order & min.~val first \\
        $m_i$ &  in order & min.~val first \\
        $r_o$ &  in order & randomly \\\hline
      \end{tabular}
    }
  }\hfill
  \subfloat[\label{tab:random} \textsc{Random} branching strategy.]{
    {\footnotesize
      \begin{tabular}{|c|>{\centering}m{2cm}|c|}
        \hline
        Variable & Var. Selection & Value Selection \\\hline 
        $c_i$ & randomly  &  randomly  \\
        $m_i$ & randomly  &  randomly \\
        $r_o$ & randomly  &  randomly  \\\hline
      \end{tabular}
    }
  }
\end{table}


\section{Evaluation}
\label{sec:evaluation}

The evaluation of \ac{DivCon} addresses four main questions:
\begin{itemize}
\item RQ1. What is the scalability of the distance measures
  in generating multiple program variants?
  Here, we evaluate which of the distance measures
  is the most appropriate for software diversification.
\item RQ2. How effective and how scalable is \ac{LNS} for code diversification? 
  Here, we investigate \ac{LNS} as an alternative approach to diversity in \ac{CP}.
\item RQ3. How does code quality relate to code diversity and what are the
  involved trade-offs?
\item RQ4. How effective is \ac{DivCon} at mitigating code-reuse attacks? 
  This question is the main application of CP-based diversification in this work.
\end{itemize}{}

\subsection{Experimental Setup}

\paragraph{Implementation.}
\ac{DivCon} is implemented as an extension of Unison~\cite{lozano_combinatorial_2019},
and is available at \url{https://github.com/romits800/divcon}.
Unison implements two backend transformations: instruction scheduling
and register allocation.
\ac{DivCon} employs Unison's solver portfolio that includes Gecode
v6.2~\cite{Gecode2020} and Chuffed v0.10.3~\cite{Chu2011} to find optimal
solutions, and Gecode v6.2 only for diversification.
The LLVM compiler~\cite{Lattner2004} is used as a front-end and
\ac{IR}-level optimizer.

\paragraph{Benchmark functions and platform.}
The evaluation uses 17 functions sampled randomly from MediaBench~\cite{Lee1997}
and SPEC CPU2006~\cite{SpecCPU2006}, two benchmark suites widely employed in
embedded and general-purpose compiler research.
The size of the functions is limited to between 10 and 30 instructions (with a
median of 20 instructions) to keep the evaluation of all methods and distance
measures feasible regardless of their computational cost.
Table~\ref{tab:benchmarks} lists the ID, application, name,
basic blocks (b), and instructions (i) of each sampled function.
The functions are compiled to MIPS32 assembly code.
MIPS32 is a popular architecture within embedded systems and the
security-critical \acl{IoT}~\cite{alaba_internet_2017}.

\paragraph{Host platform.}
All experiments run on
an Intel%
\textsuperscript{\textregistered}%
Core\texttrademark i9-9920X processor at 3.50GHz with 64GB of RAM
running Debian GNU/Linux 10 (buster).
Each of the experiments runs for 20 random seeds.
The results
show the mean value and the standard deviation
from these experiments.
The available virtual memory for each
of the executions is 10GB. 
The experiments for different random seeds run in parallel (5 seeds at a time), with
two unique cores 
available for every seed for overheating reasons.
\ac{DivCon} runs as a sequential program.

\paragraph{Algorithm Configuration.}
The experiments focus on speed optimization and aim to generate 200
variants within a timeout.
Parameter $h$ in Algorithm~\ref{alg:search} is set to one because even
small distance between variants is able to break gadgets (see Figure~\ref{fig:mips_example}).
\ac{LNS} uses restart-based search with a limit of 500 failures, and a relax
rate of 70\%.
%
The \textit{relax rate} is the probability that \ac{LNS} destroys
a variable at every restart,
which affects the distance between two subsequent solutions.
A higher relax rate increases diversity but requires more solving effort.
We have found experimentally
that 70\% is an adequate balance between the two.
All experiments are available at \url{https://github.com/romits800/divcon\_experiments}.

\subsection{RQ1. Scalability of the Distance Measures}
\label{ssec:scale_dist}

The ability to generate a large number of variants
is paramount
for software diversification.
This section compares the distance measures introduced in Section~\ref{ssec:distances} with regards to scalability.

\begin{figure*}[t]
  \centering
\begin{minipage}[t]{0.55\textwidth}
\begin{scriptsize}
   \setlength\tabcolsep{2pt}
   \begin{longtable}{|l|l|l|l|l|}
\caption{\label{tab:benchmarks}{Benchmark functions}}\\
\hline
{ID}&{app}&{function name}&{b}&{i}\\
\hline
b1&sphinx3&ptmr\_init&1&10
\\
b2&gcc&ceil\_log2&1&14
\\
b3&mesa&glIndexd&1&14
\\
b4&h264ref&symbol2uvlc&1&15
\\
b5&gobmk&autohelperowl\_defen..&1&23
\\
b6&mesa&glVertex2i&1&23
\\
b7&hmmer&AllocFancyAli&1&25
\\
b8&gobmk&autohelperowl\_vital..&1&27
\\
b9&gobmk&autohelperpat1088&1&29
\\
b10&gobmk&autohelperowl\_attac..&1&30
\\
b11&gobmk&get\_last\_player&3&13
\\
b12&h264ref&UpdateRandomAccess&3&16
\\
b13&gcc&xexit&3&17
\\
b14&gcc&unsigned\_condition&3&24
\\
b15&sphinx3&glist\_tail&4&10
\\
b16&gcc&get\_frame\_alias\_set&5&20
\\
b17&gcc&parms\_set&5&25
\\
\hline
\end{longtable}

\end{scriptsize}
\end{minipage}\hfill
\begin{minipage}[t]{0.40\textwidth}
\begin{scriptsize}
   \setlength\tabcolsep{2pt}
   \begin{longtable}{|l|c|c|c|c|}
\caption{\label{tab:distances}{Scalability of $\delta_{HD}, \delta_{LD}$}}\\
\hline
\multirow{2}{*}{ID}&\multicolumn{2}{c|}{$\delta_{HD}$}&\multicolumn{2}{c|}{$\delta_{LD}$}\\
\cline{2-5}
&$t(s)$&num&$t(s)$&num\\
\hline
b1&0.1$\pm$0.2 & 26&131.2$\pm$131.4 & 26
\\
b2&1.0$\pm$0.1 & 200 &- & 68
\\
b3&1.1$\pm$0.1 & 200 &- & 58
\\
b4&0.7$\pm$0.0 & 200 &- & 73
\\
b5&2.3$\pm$0.3 & 200 &- & 38
\\
b6&2.5$\pm$0.2 & 200 &- & 35
\\
b7&2.0$\pm$0.3 & 200 &- & 37
\\
b8&3.8$\pm$0.8 & 200 &- & 35
\\
b9&4.0$\pm$0.6 & 200 &- & 28
\\
b10&4.5$\pm$0.7 & 200 &- & 27
\\
b11&1.3$\pm$0.1 & 200 &- & 56
\\
b12&1.1$\pm$0.2 & 200 &- & 47
\\
b13&0.8$\pm$0.1 & 200 &- & 91
\\
b14&1.8$\pm$0.3 & 200 &- & 27
\\
b15&1.7$\pm$0.2 & 200 &- & 60
\\
b16&2.7$\pm$0.4 & 200 &- & 31
\\
b17&1.6$\pm$0.2 & 200 &- & 35
\\
\hline
\end{longtable}

\end{scriptsize}
\end{minipage}
\end{figure*}

Table~\ref{tab:distances} presents the results of the distance evaluation, where
a time limit of 10 minutes and optimality gap of $p=10\%$ are used.
For each distance measure ($\delta_{HD}$ and $\delta_{LD}$) the table
shows the diversification time $t$, in seconds (or ``-'' if
the algorithm is not able to generate 200 variants) and the number of generated
variants $num$ within the time limit.

The results show that for $\delta_{HD}$, \ac{DivCon} is able to generate
200 variants for all benchmarks except \emph{b1}, which
has exactly 26 variants. The diversification time for $\delta_{HD}$ is
less than 5 seconds for all benchmarks.
Distance $\delta_{LD}$, on the other hand, is not able to generate 200 variants
for any of the benchmarks within the time limit.
This poor scalability of $\delta_{LD}$ is due to
the quadratic complexity of its implementation~\cite{wagner1974string}, whereas
\ac{HD} can be implemented linearly.  Consequently, the rest of the evaluation
uses $\delta_{HD}$.

\subsection{RQ2. Scalability and Diversification Effectiveness of \ac{LNS}}
\label{ssec:evallns}

This section evaluates the diversification effectiveness and scalability of
\ac{LNS} compared to incremental \textsc{MaxDiverse$k$Set} (where the
first solution is found randomly and the
maximization step uses the branching strategy from
Table~\ref{tab:cloriginal}) and \ac{RS} (which uses the branching strategy from
Table~\ref{tab:random}). 

To measure the diversification effectiveness of these methods,
the evaluation uses the relative pairwise distance
of the solutions.
Given a set of solutions $S$ and a distance measure $\delta$,
the pairwise distance $d$
of the variants in $S$ is
$d(\delta, S) = \left. \sum_{i=0}^{|S|} \sum_{j>i}^{|S|} \delta(s_i, s_j) \, / \, {\binom{|S|}{2}}\right.$.
The \emph{larger} this distance, the more
diverse the solutions are, and thus, diversification is more effective. 
%
Table~\ref{tab:dist_max_rs_lns} shows the pairwise distance $d$
and diversification time $t$ for each benchmark and method,
where the experiment uses a time limit
of 30 minutes and optimality gap of $p=10\%$.
The best values of $d$ (larger) and $t$ (lower) are marked in \textbf{bold} for the completed
experiments, whereas incomplete experiments are highlighted in \textit{italic}
and their number of variants in parenthesis.

\begin{scriptsize}
  \setlength\tabcolsep{3pt}
  \begin{longtable}{|l|c|c|c|c|c|c|}
\caption{\label{tab:dist_max_rs_lns} Distance and Scalability of \ac{LNS} with \ac{RS} and \textsc{MaxDiverse$k$Set}}\\
\hline
\multirow{2}{*}{ID}&\multicolumn{2}{c|}{\textsc{MaxDiverse$k$Set}}&\multicolumn{2}{c|}{{RS}}&\multicolumn{2}{c|}{LNS (0.7)}\\
\cline{2-7}
&$d$&$t(s)$&$d$&$t(s)$&$d$&$t(s)$\\
\hline
b1&\textit{4.1$\pm$0.0} & 0.2$\pm$0.0 (26)&\textit{4.1$\pm$0.0} & \textbf{0.0$\pm$0.0 (26)}&\textit{4.1$\pm$0.0} & 0.1$\pm$0.2 (26)
\\
b2&\textbf{10.8$\pm$0.0} & 761.8$\pm$10.1&6.4$\pm$0.2 & \textbf{0.6$\pm$0.1}&8.6$\pm$0.6 & 1.0$\pm$0.1
\\
b3&\textit{14.6$\pm$0.0} & - (21)&5.8$\pm$0.1 & \textbf{0.6$\pm$0.1}&\textbf{10.8$\pm$0.8} & 1.0$\pm$0.1
\\
b4&\textit{14.4$\pm$0.0} & - (19)&4.3$\pm$0.1 & \textbf{0.2$\pm$0.0}&\textbf{12.1$\pm$0.3} & 0.6$\pm$0.0
\\
b5&\textit{22.0$\pm$0.0} & - (2)&4.3$\pm$0.3 & \textbf{0.5$\pm$0.0}&\textbf{16.1$\pm$1.1} & 2.2$\pm$0.3
\\
b6&\textit{22.9$\pm$0.4} & - (2)&5.3$\pm$0.0 & \textbf{1.0$\pm$0.1}&\textbf{16.4$\pm$0.6} & 2.4$\pm$0.2
\\
b7&\textit{24.9$\pm$0.1} & - (6)&4.5$\pm$0.2 & \textbf{0.4$\pm$0.0}&\textbf{18.1$\pm$1.2} & 1.9$\pm$0.3
\\
b8&\textit{24.8$\pm$0.4} & - (2)&6.5$\pm$0.2 & \textbf{3.5$\pm$0.5}&\textbf{17.2$\pm$0.9} & 3.8$\pm$0.8
\\
b9&\textit{26.0$\pm$0.0} & - (2)&4.2$\pm$0.3 & \textbf{0.4$\pm$0.0}&\textbf{19.8$\pm$0.7} & 3.9$\pm$0.6
\\
b10&\textit{28.0$\pm$0.0} & - (2)&6.0$\pm$0.0 & 5.3$\pm$1.0&\textbf{20.1$\pm$1.1} & \textbf{4.5$\pm$0.7}
\\
b11&\textbf{13.8$\pm$0.0} & 356.9$\pm$8.2&5.3$\pm$0.1 & \textbf{0.2$\pm$0.0}&10.1$\pm$1.0 & 1.2$\pm$0.1
\\
b12&\textit{21.5$\pm$0.1} & - (5)&6.4$\pm$0.9 & \textbf{0.2$\pm$0.0}&\textbf{14.9$\pm$1.0} & 1.0$\pm$0.2
\\
b13&\textit{17.4$\pm$0.0} & - (122)&6.7$\pm$0.0 & 0.9$\pm$0.1&\textbf{12.0$\pm$0.9} & \textbf{0.7$\pm$0.1}
\\
b14&\textit{30.1$\pm$0.0} & - (20)&7.5$\pm$0.2 & \textbf{0.2$\pm$0.0}&\textbf{24.9$\pm$0.7} & 1.8$\pm$0.3
\\
b15&- & -&2.6$\pm$0.3 & \textbf{0.1$\pm$0.0}&\textbf{20.2$\pm$0.5} & 1.6$\pm$0.2
\\
b16&- & -&5.6$\pm$0.4 & \textbf{0.3$\pm$0.0}&\textbf{21.3$\pm$0.8} & 2.6$\pm$0.4
\\
b17&- & -&\textit{2.9$\pm$0.1} & - (91)&\textbf{28.1$\pm$1.5} & \textbf{1.6$\pm$0.2}
\\
\hline
\end{longtable}

\end{scriptsize}

The scalability results ($t(s)$) show
that \ac{RS} and \ac{LNS} are scalable (generate the maximum 200 variants for almost all benchmarks),
whereas \textsc{MaxDiverse$k$Set} scales poorly (cannot generate 200 variants for any benchmark but \textit{b2} and \textit{b11}).
Both \textit{b2} and \textit{b11}
have a small search space (few, highly interdependent instructions), which leads to restricted diversity but facilitates solving.
For \textit{b1}, 
all instructions are interdependent on each other, which
forces a linear schedule and results in only 26 possible variants (given $p=10\%$).
On the other end,  \textsc{MaxDiverse$k$Set}
is not able to find any variants for \textit{b15}, \textit{b16},
and \textit{b17}.
These benchmarks have many basic blocks
resulting in a more complex objective function.
For the largest benchmark (\textit{b17}), only \ac{LNS} is able to scale  up to 200 solutions.
\ac{LNS} is generally slower than \ac{RS}, but for both \ac{LNS} and \ac{RS} all
benchmarks have a diversification time less than six seconds.

The diversity results ($d$) show that \ac{LNS} is more effective at diversifying than
\ac{RS}. The improvement of \ac{LNS} over \ac{RS} ranges from $35\%$ (for
\textit{b2}) to $675\%$ (for \textit{b15}).
In the two cases where \textsc{MaxDiverse$k$Set} terminates (benchmarks
\textit{b2} and \textit{b11}), it generates the most diverse code, as can be
expected.

In summary, \ac{LNS} offers an attractive balance between scalability and
diversification effectiveness: it is close in scalability to, and sometimes
improves, the overly fastest method (\ac{RS}), but it is significantly and consistently more
effective at diversifying code.

\subsection{RQ3. Trade-off Between Code Quality and Diversity}
\label{ssec:gap_eval}

A key advantage of using a
\ac{CP}-based compiler approach for
software diversity is the ability to control the
quality of the generated solutions.
This ability enables control over the
relation between the quality of each individual solution and the diversity of the entire pool of solutions.
Insisting in optimality limits
the number of possible diversified variants
and their pairwise distance,
whereas relaxing optimality allows
higher diversity.


Table~\ref{tab:agaps} shows the pairwise distance $d$
(defined in Section~\ref{ssec:evallns}),
and the number of generated variants $num$,
for all benchmarks and different values of the optimality gap $p\in$\{0\%, 5\%, 10\%, 20\%\}.
\ac{LNS} is used with a time limit of 10 minutes.
The best values of $d$ are marked in \textbf{bold}.

\begin{scriptsize}
  \setlength\tabcolsep{3pt}
  \begin{longtable}{|l|c|c|c|c|c|c|c|c|}
\caption{\label{tab:agaps} Solution diversity for different optimality gap values}\\
\hline
\multirow{2}{*}{ID}&\multicolumn{2}{c|}{0\%}&\multicolumn{2}{c|}{5\%}&\multicolumn{2}{c|}{10\%}&\multicolumn{2}{c|}{20\%}\\
\cline{2-9}
&$d$&num&$d$&num&$d$&num&$d$&num\\
\hline
b1&- & -&- & -&\textit{4.1$\pm$0.0} & 26&\textbf{6.5$\pm$0.1} & 200
\\
b2&\textit{3.5$\pm$0.0} & 9&6.7$\pm$0.4 & 200&8.6$\pm$0.6 & 200&\textbf{10.0$\pm$0.8} & 200
\\
b3&7.0$\pm$0.1 & 200&9.4$\pm$0.5 & 200&10.8$\pm$0.8 & 200&\textbf{14.8$\pm$1.0} & 200
\\
b4&7.8$\pm$0.2 & 200&10.1$\pm$0.3 & 200&12.1$\pm$0.3 & 200&\textbf{14.0$\pm$0.2} & 200
\\
b5&8.4$\pm$0.1 & 200&11.9$\pm$0.7 & 200&16.1$\pm$1.1 & 200&\textbf{19.7$\pm$0.6} & 200
\\
b6&10.8$\pm$0.1 & 200&14.7$\pm$0.4 & 200&16.4$\pm$0.6 & 200&\textbf{20.9$\pm$0.8} & 200
\\
b7&11.3$\pm$0.3 & 200&13.8$\pm$0.7 & 200&18.1$\pm$1.2 & 200&\textbf{22.8$\pm$1.1} & 200
\\
b8&11.0$\pm$0.1 & 200&13.6$\pm$0.6 & 200&17.2$\pm$0.9 & 200&\textbf{22.4$\pm$1.1} & 200
\\
b9&12.7$\pm$0.1 & 200&17.7$\pm$0.8 & 200&19.8$\pm$0.7 & 200&\textbf{24.4$\pm$0.6} & 200
\\
b10&13.7$\pm$0.1 & 200&18.1$\pm$0.9 & 200&20.1$\pm$1.1 & 200&\textbf{26.3$\pm$0.6} & 200
\\
b11&\textit{2.0$\pm$0.0} & 4&6.6$\pm$0.1 & 200&10.1$\pm$1.0 & 200&\textbf{14.2$\pm$0.9} & 200
\\
b12&\textit{3.8$\pm$0.0} & 10&10.3$\pm$1.2 & 200&14.9$\pm$1.0 & 200&\textbf{19.8$\pm$1.0} & 200
\\
b13&\textit{2.1$\pm$1.3} & 4&10.1$\pm$0.9 & 200&12.0$\pm$0.9 & 200&\textbf{15.7$\pm$1.2} & 200
\\
b14&\textit{3.6$\pm$0.0} & 24&21.0$\pm$0.6 & 200&24.9$\pm$0.7 & 200&\textbf{29.0$\pm$0.5} & 200
\\
b15&\textit{2.4$\pm$0.0} & 8&15.6$\pm$0.6 & 200&20.2$\pm$0.5 & 200&\textbf{23.5$\pm$1.4} & 200
\\
b16&\textit{4.1$\pm$0.0} & 44&15.1$\pm$1.1 & 200&21.3$\pm$0.8 & 200&\textbf{30.7$\pm$0.9} & 200
\\
b17&7.5$\pm$0.2 & 200&20.3$\pm$1.4 & 200&28.1$\pm$1.5 & 200&\textbf{38.4$\pm$0.9} & 200
\\
\hline
\end{longtable}

\end{scriptsize}

The first interesting observation is that even with no degradation of quality ($p=0\%$),
\ac{DivCon} is able to generate a large number of variants
for a significant fraction of the benchmarks.
%
These include functions with a relatively large solution space, typically with a
few large basic blocks where instructions are relatively independent of each
other (\textit{b3-b10} and \textit{b17}).
On the other hand, benchmarks with small basic blocks and many instruction
dependencies (\textit{b1}, \textit{b2}, and \textit{b11-b16}) provide fewer
options for diversification, which results in a limited number of optimal
variants.

Second, we observe that as soon as we slightly relax the
constraint over optimality ($p=5\%$),
diversity radiates and \ac{DivCon} generates 200 variants for all benchmarks
except \emph{b1}.
Then, the more we increase the optimality gap,
the larger the diversification space grows
and the distance between the variants increases.
Table~\ref{tab:agaps} illustrates one of the key contributions of \ac{DivCon}:
the ability to explore the trade-off between optimal solutions and highly diverse solutions.

In summary, depending on the characteristics of the compiled code, it is
possible to generate a large number of variants without sacrificing optimality,
and the code quality can be adjusted to further improve diversity if required by
the targeted application.

\subsection{RQ4. Code-Reuse Mitigation Effectiveness}
\label{ssec:sec_eval}

Software Diversity has various applications in security, including
mitigating code-reuse attacks.
To measure the level of mitigation that \ac{DivCon} achieves, we assess the
gadget survival rate $srate(s_i,s_j)$ between two variants $s_i,s_j \in S$,
where $S$ is the set of generated variants.  This metric determines how many of
the gadgets of variant $s_i$ appear at the same position on the other variant
$s_j$, that is $srate(s_i,s_j) = |gad(s_i) - gad(s_j)| \, / \, |gad(s_i)|$,
where $gad(s_i)$ are the gadgets in solution $s_i$.  The procedure for computing
$srate(s_i,s_j)$ is as follows: 1) run ROPgadget~\cite{ROPGadget2020} to find
the set of gadgets $gad(s_i)$ in solution $s_i$, and 2) for every $g\in
gad(s_i)$, check whether there exists a gadget identical to $g$ at the same
address of $s_j$.  This comparison is syntactic after
removing all \texttt{nop} instructions.

This section compares the $srate$
for all permutations
of pairs in $S$,
for all benchmarks, and
for different
values of the optimality gap using a time limit of 10 minutes.
Low $srate$ corresponds to higher mitigation effectiveness
because code-reuse attacks based on gadgets in one variant
have lower chances of locating the same gadgets in the other variants
(see Figure~\ref{fig:mips_example}).

Table~\ref{tbl:optimal} summarizes the gadget survival distribution
for all benchmarks and different values of the
optimality gap (0\%, 5\%, 10\%, and 20\%).
Due to its skewness, the distribution of $srate$ is represented as a histogram
with four buckets (0\%, (0\%, 10\%], (10\%,40\%], and (40\%, 100\%]) rather than
      summarized using common statistical measures.
Here the best is a $srate(s_i,s_j)$ of 0\%, which means that $s_j$ does not
contain any gadgets that exist in $s_i$, whereas
a $srate(s_i,s_j)$ in range (40\%,100\%] means that $s_j$  shares more than 40\% of the gadgets of $s_i$.
The values in \textbf{bold} correspond to the mode(s) of the histogram.

First, we notice that \ac{DivCon} can generate some pairs of variants that share no gadget, even without relaxing the constraint of optimality ($p = 0\%$). This indicates that the pareto front of optimal code naturally includes software diversity that is good for security. Second, the results show that
this effectiveness can be further increased by relaxing the constraint on code quality, with diminishing returns beyond $p = 10\%$.
For $p = 0\%$, there are 10 benchmarks dominated by a $0\%$ survival rate,
whereas there are 7 benchmarks dominated by a weak $10\%-40\%$-survival rate.
The latter are still considered vulnerable to code-reuse attacks.
However, increasing the optimality gap to just $p = 5\%$ makes $0\%$ survival
rate the dominating bucket for all benchmarks, and further increasing the gap to
$10\%$ and $20\%$ increases significantly the number of pairs where no single
gadget is shared.
For example, at $p = 10\%$ the rate of pairs that do not share any gadgets
ranges from 63\% (\emph{b14}) to 99\% (\emph{b12}).

\begin{table}[t]
  \caption{\label{tbl:optimal}
    Gadget survival rate for different optimality gap values}
  \setlength\tabcolsep{1pt}
 \begin{filecontents*}{tables/hist_gaps_output0.7hamming.csv}
,\tiny{$=$0},\tiny{$\le$10},\tiny{$\le$40},\tiny{$\le$100},\tiny{num},\tiny{$=$0},\tiny{$\le$10},\tiny{$\le$40},\tiny{$\le$100},\tiny{num},\tiny{$=$0},\tiny{$\le$10},\tiny{$\le$40},\tiny{$\le$100},\tiny{num},\tiny{$=$0},\tiny{$\le$10},\tiny{$\le$40},\tiny{$\le$100},\tiny{num}
b1,-,-,-,-,-,-,-,-,-,-,\textbf{84},3,3,10,26,\textbf{94},4,2,1,200
b2,-,-,\textbf{69},31,9,\textbf{60},12,23,4,200,\textbf{76},11,12,1,200,\textbf{81},9,10,-,200
b3,\textbf{66},15,18,1,200,\textbf{71},14,15,1,200,\textbf{73},13,13,1,200,\textbf{77},14,9,-,200
b4,\textbf{94},6,-,-,200,\textbf{96},4,-,-,200,\textbf{96},4,-,-,200,\textbf{98},2,-,-,200
b5,\textbf{90},1,9,-,200,\textbf{93},2,5,-,200,\textbf{95},2,3,-,200,\textbf{95},3,2,-,200
b6,\textbf{88},5,7,1,200,\textbf{89},5,6,-,200,\textbf{90},4,6,-,200,\textbf{91},4,5,-,200
b7,\textbf{48},1,\textbf{48},3,200,\textbf{74},5,21,1,200,\textbf{83},6,11,-,200,\textbf{89},6,5,-,200
b8,\textbf{46},-,\textbf{51},3,200,\textbf{57},4,36,2,200,\textbf{74},3,21,1,200,\textbf{81},4,14,1,200
b9,42,-,\textbf{56},2,200,\textbf{66},9,24,1,200,\textbf{73},8,18,-,200,\textbf{83},7,9,-,200
b10,\textbf{47},-,\textbf{50},3,200,\textbf{65},2,30,2,200,\textbf{73},4,22,1,200,\textbf{82},5,13,1,200
b11,38,-,\textbf{61},1,4,\textbf{66},3,31,-,200,\textbf{68},9,23,-,200,\textbf{83},7,10,-,200
b12,\textbf{94},-,5,1,10,\textbf{99},1,-,-,200,\textbf{99},-,-,-,200,\textbf{99},1,-,-,200
b13,\textbf{43},9,34,14,4,\textbf{69},20,11,-,194,\textbf{69},21,10,-,200,\textbf{71},19,10,-,200
b14,-,-,\textbf{78},22,24,\textbf{60},23,17,-,200,\textbf{63},22,15,-,200,\textbf{70},19,11,-,200
b15,41,\textbf{53},5,-,8,\textbf{97},2,1,-,200,\textbf{98},1,1,-,200,\textbf{98},1,1,-,200
b16,\textbf{64},28,6,-,44,\textbf{76},21,2,-,200,\textbf{82},17,1,-,200,\textbf{90},9,1,-,200
b17,33,\textbf{66},1,-,200,\textbf{61},39,-,-,200,\textbf{75},25,-,-,200,\textbf{87},13,-,-,200
\end{filecontents*}
\pgfplotstabletypeset[
   col sep=comma,
   string type,
   columns={0,1,2,3,4,5,6,7,8,9,10,11,12,13,14,15,16,17,18,19,20},
   columns/0/.style= {column type={|M{14pt}|}},
   columns/1/.style= {column type={M{15pt}}},
   columns/2/.style= {column type={M{14pt}}},
   columns/3/.style= {column type={M{14pt}}},
   columns/4/.style= {column type={M{15pt}}},
   columns/5/.style= {column type={|M{14pt}|}},
   columns/6/.style={column type={M{15pt}}},
   columns/7/.style={column type={M{14pt}}},
   columns/8/.style={column type={M{14pt}}},
   columns/9/.style={column type={M{15pt}}},
   columns/10/.style={column type={|M{14pt}|}},
   columns/11/.style={column type={M{15pt}}},
   columns/12/.style={column type={M{14pt}}},
   columns/13/.style={column type={M{14pt}}},
   columns/14/.style={column type={M{15pt}}},
   columns/15/.style={column type={|M{14pt}|}},
   columns/16/.style={column type={M{15pt}}},
   columns/17/.style={column type={M{14pt}}},
   columns/18/.style={column type={M{14pt}}},
   columns/19/.style={column type={M{15pt}}},
   columns/20/.style={column type={|M{14pt}|}},
   header=false,
   every first row/.style={after row=\hline},
   every head row/.style={before row=\hline\multirow{2}{*}{ID} & \multicolumn{5}{c|}{\scriptsize{0\%}} & \multicolumn{5}{c|}{\scriptsize{5\%}} & \multicolumn{5}{c|}{\scriptsize{10\%}} & \multicolumn{5}{c|}{\scriptsize{20\%}}\\\cline{2-21},output empty row},
   every last row/.style={after row=\hline},
   font=\scriptsize,
 ]{tables/hist_gaps_output0.7hamming.csv}
\end{table}
   
Related approaches (discussed in Section~\ref{sec:rel}) report the
\emph{average} $srate$ across all pairs for different benchmark sets.
Pappas \emph{et al.}'s zero-cost approach~\cite{pappas_smashing_2012} achieves
an average $srate$ between $74\%-83\%$ without code degradation, comparable to
\ac{DivCon}'s $41\%-99\%$ at $p=0\%$.
Homescu \emph{et al.}'s statistical approach~\cite{homescu_profile-guided_2013}
reports an average $srate$ between $82\%-100\%$ with a code degradation of less
than $5\%$, comparable to \ac{DivCon}'s $83\%-100\%$ at $p=5\%$.
Both approaches report results on larger code bases that exhibit more
opportunities for diversification.
We expect that \ac{DivCon} would achieve higher overall survival rates on these
code bases compared to the benchmarks used in this paper.

\section{Related Work}
\label{sec:rel}

There are many approaches to software diversification
against cyberattacks.
The majority
apply randomized transformations 
at different stages of the software development, while
a few exceptions use search-based techniques 
\cite{larsen_sok_2014}.
This section focuses on quality-aware software diversification approaches.

\textit{Superdiversifier}~\cite{jacob_superdiversifier_2008} is a search-based approach
for software diversification against cyberattacks.
Given an initial instruction sequence, the algorithm generates a random
combination of the available instructions
and performs a verification test to quickly reject
non equivalent instruction sequences.
For each non-rejected sequence,
the algorithm checks semantic equivalence between the original
and the generated instruction sequences
using a SAT solver.
Superdiversifier affects the code execution time and size
by controlling the length of the generated
sequence.
Along the same lines,
Lundquist \emph{et al.}\  \cite{lundquist_searching_2016,lundquist2019relational}
use program synthesis
for generating program variants against
cyberattacks, but no results are available yet.
In comparison, \ac{DivCon} uses a combinatorial
compiler backend that
measures the code quality using a more accurate
cost model that also
considers
other aspects, such as execution frequencies.

Most diversification approaches
use randomized  transformations to
generate  multiple program variants \cite{larsen_sok_2014}.
Unlike \ac{DivCon}, the majority of these approaches
do not control the quality of the generated variants
during diversification but rather evaluate it afterwards
\cite{davi2013gadge,wang_composite_2017,koo_compiler-assisted_2018,homescu_large-scale_2017,braden_leakage-resilient_2016,crane_readactor_2015}.
However, there are a few approaches that control
the code quality
during randomization.

Some compiler-based diversification approaches
restrict the set of program transformations to control
the quality of the generated code \cite{crane_readactor_2015,pappas_smashing_2012}.
For example, Pappas \emph{et al.}\  \cite{pappas_smashing_2012}
perform software diversification at the binary level and apply
three zero-cost transformations: register randomization,
instruction schedule randomization, and function
shuffling.
In contrast, \ac{DivCon}'s combinatorial approach allows it to control the
aggressiveness and potential cost of its transformations: a cost overhead limit
of 0\% forces \ac{DivCon} to apply only zero-cost transformations; a larger
limit allows \ac{DivCon} to apply more aggressive transformations, potentially
leading to higher diversity.

Homescu \emph{et al.}\  \cite{homescu_profile-guided_2013}
perform only garbage (\texttt{nop}) insertion, and use a profile-guided approach
to reduce the overhead.
To do this, they control the \texttt{nop} insertion probability
based on the execution frequency of different code sections.
In contrast, \ac{DivCon}'s cost model captures different execution frequencies,
which allows it to perform more aggressive
transformations in non-critical code sections.

\section{Conclusion and Future Work}
\label{sec:conclusion_fw}
This paper  introduces \ac{DivCon}, a \ac{CP} approach to compiler-based,
quality-aware software diversification against code-reuse attacks. Our
experiments show that \ac{LNS} is a promising technique for a
\ac{CP}-based exploration of the space of diverse program, with a fine-grained control on the trade-off between code
quality and diversity. In particular, we show that the set of optimal solutions naturally contains a set of diverse solutions, which increases significantly when relaxing the constraint of optimality. Our experiments demonstrate that the diverse solutions generated by DivCon are effective to mitigate code-reuse attacks.

Future work includes investigating different distance measures to further
reduce the gadget survival rate, improving the overall scalability of
\ac{DivCon} in the face of larger programs and larger values of parameter $k$,
and examining the effectiveness of
\ac{DivCon} against an actual code-reuse exploit.

\subsubsection{Acknowledgments.}
We would like to give a special acknowledgment
to Christian Schulte,
for his critical contribution at the early stages of this work.
Although no longer with us, Christian continues to inspire
his students and colleagues
with his
lively character, enthusiasm,
deep knowledge and understanding.
We would also like to thank Linnea Ingmar and the anonymous reviewers
for their useful feedback, and
Oscar Eriksson for proof reading.

\bibliographystyle{splncs04}
\bibliography{bibliography}

\end{document}